\documentclass[10pt,showpacs,twocolumn]{revtex4}
\usepackage{graphicx}
\usepackage{amsmath}
\usepackage{amssymb}
\usepackage{multirow}
\usepackage{color}
\usepackage{}

\begin{document}
\title{Time-dependent population imaging for solid high harmonic generation}
 \author{Xi Liu,$^{1}$ Xiaosong Zhu,$^{1}$ \footnote{zhuxiaosong@hust.edu.cn} Pengfei Lan,$^{1}$ Xiaofan Zhang,$^{1}$ Dian Wang,$^{1}$ Qingbin Zhang,$^{1}$   Peixiang Lu$^{1,2}$ \footnote{lupeixiang@mail.hust.edu.cn}}

\affiliation{$^1$  School of Physics and Wuhan National Laboratory for Optoelectronics, Huazhong University of
Science and Technology, Wuhan 430074, China\\
$^2$ Laboratory of Optical Information Technology, Wuhan Institute of Technology, Wuhan 430205, China}
\date{\today}

\begin{abstract}
We propose an intuitive method, called time-dependent population imaging (TDPI), to map the dynamical processes of high harmonic generation (HHG) in solids by solving the time-dependent Schr\"{o}dinger equation (TDSE). It is shown that the real-time dynamical characteristics of HHG in solids, such as the instantaneous photon energies of emitted harmonics, can be read directly from the energy-resolved population oscillations of electrons in the TDPIs. Meanwhile, the short and long trajectories of solid HHG are illustrated clearly from TDPI. By using the TDPI, we also investigate the effects of carrier-envelope phase (CEP) in few-cycle pulses and intuitively demonstrate the HHG dynamics driven by two-color fields. Our results show that the TDPI provides a powerful tool to study the ultrafast dynamics in strong fields for various laser-solid configurations and  gain an insight into HHG processes in solids.
\end{abstract}  \maketitle

\noindent \section{Introduction}
Many interesting strong-field phenomena have been revealed when atoms and molecules interact with intense laser fields \cite{Krausz,Corkum,Itatani,Zhouyueming}. One of the most fascinating phenomena is high harmonic generation (HHG) \cite{Krause1992,Schafer1993}. The HHG from gas phase has been studied widely over the past several decades \cite{Corkum_2,Lewenstein,HHG_lu}. Recently, the experimental observation of HHG from bulk solids has attracted extensive attentions in the field of attosecond science \cite{Ghimire,Ghimire-JPB}. Apart from having the traditional advantages of gas HHG, the solid HHG also has the great potential superiority to achieve higher conversion efficiency due to the high density of solid targets \cite{Ghimire-PRA}. This property make it a competitive alternative of obtaining the table-top extreme ultraviolet (XUV) light source \cite{Luu,Cavalieri}. In addition, solid HHG provides a useful tool to probe the energy band structures of crystals \cite{Wang}, and even to image the orbitals of solids \cite{Ndabashimiye}. For example, Vampa {\it et al.} \cite{Vampa_AO} reconstructed the energy bands of ZnO crystal based on HHG method. The solid HHG has opened up a new frontier to study the attosecond electron dynamics in condensed matter \cite{Schultze}.

For the solid HHG, the particular characteristics distinguishing from gas HHG essentially stem from the periodicity and high density of the crystal. At present, the driving wavelengths for solids are centered in mid-infrared (MIR) \cite{Ghimire,Ghimire-PRL} and terahertz \cite{Zaks,Hohenleutner} regions. With such long wavelengths, the laser intensity can be moderate and thus lower than the damage threshold. The harmonic spectra from solids exhibit evident multiple plateaus \cite{Wu2016,Wu2015} and extend well beyond the atomic limit. The cutoff energy of the solid HHG shows a linear dependence on field strength \cite{Ghimire,Ghimire-JPB}, unlike the quadratic dependence relation in gas HHG. The ellipticity dependence of HHG in solids is complicated. The experimental result shows that ellipticity dependence of HHG from ZnO crystal is much weaker than that in gas harmonics \cite{Ghimire}, whereas harmonics from the rare-gas solids (RGS) show a strong ellipticity dependence as from gaseous atoms \cite{Ndabashimiye}.

The mechanism of HHG in solids has been a topic of intense debate \cite{Higuchi,Hawkins2015,Kemper}. In most works, it is considered that the HHG in solids originates from two distinct contributions: an intraband current in the individual bands and an interband current involving the transitions between the valence and conduction bands. Theoretical analyses show that the interband current dominates the HHG for MIR driver pulses \cite{McDonald,Vampa-PRB, Guan}. Vampa {\it et al.} \cite{Vampa-Nature, Vampa-PRL} proposed an electron-hole recollision model to describe the mechanism, where electrons in conduction bands recombine with associated holes in the valence band. Meanwhile, Wu {\it et al.} \cite{Wu2016,Wu2015} suggested that the primary plateau originates from transitions from the first conduction to the valence band and the latter plateaus are due to transitions from higher-lying conduction bands. However, when driving wavelengths are extended toward the terahertz regime, the intraband current caused by laser-driven Bloch oscillations becomes dominant for the HHG processes \cite{Schubert,Hawkins,Golde,Vampa-PRB}. Currently, although some theoretical models can well explain the solid HHG, an intuitive method to describe the picture of HHG in solids is still an urgent demand.

In this work, we propose an intuitive picture, named time-dependent population imaging (TDPI), to reveal the HHG process in solids. In TDPIs, the real-time dynamics of HHG are mapped intuitively via the population oscillations of electrons at different energy bands. The features of HHG can be directly decoded from TDPIs. For example, the photon energies of real-time harmonic emissions can be read from the instantaneous energy differences between the oscillating electrons in different bands. In particular, the cutoff energies are obtained according to the maximum energy differences shown in TDPIs. The short and long trajectories of solid HHG can be distinguished clearly in TDPI picture. The carrier--envelope phase (CEP) effects in few-cycle pulses and HHG processes in two--color fields are also demonstrated by using the TDPIs. The TDPI approach can be used to visualize the solid HHG dynamics, and advance our understanding of strong-field and attosecond physics in solids.

This paper is organized as follows. In Sec.\ \ref{model}, we describe the theoretical model and numerical method in our simulations. In Sec.\ \ref{TDPI}, we introduce the TDPIs and discuss the start and cutoff energies of solid HHG with TDPIs. In Sec.\ \ref{trajectories}, the concepts of short and long trajectories for solid HHG are established based on TDPIs. In Secs.\ \ref{CEP}, the TDPI is used to explain the CEP effect in solid HHG. In Secs.\ \ref{twocolor}, TDPI is used to reveal the HHG processes driven by two-color fields, and the effect of the relative phase between the two components on the HHG dynamics is discussed. We summarize our works in Sec.\ \ref{conclusion}.

\noindent \section{THEORETICAL MODEL} \label{model}
In our simulation, we describe the laser--crystal interaction with a one-dimensional single-active electron system. The laser field is polarized along crystal axis. Since the wavelengths we are interested in are much larger than the lattice constant, dipole approximation is valid and has been employed in our calculation. In the length gauge, the time-dependent Hamiltonian reads as (atomic units are used throughout this paper unless otherwise stated)
\begin{eqnarray} \label{H_t}
\hat{H}(t) = \hat{H}_0+xF(t),
\end{eqnarray}
where $\hat{H}_0$ is the field-free Hamiltonian and $F(x)$ is the electric field of driving laser. $\hat{H}_0$ is written as $\hat{H}_0= \hat{p}^2/2+V(x)$, where $\hat{p}$ is the momentum operator and $V(x)$ is the periodic lattice potential. Herein, we use the Mathieu-type potential $V(x)=-V_0[1+\cos(2\pi x/a_0)]$ with $V_0=0.37 \ \rm{a.u.}$ and the lattice constant $a_0=8 \ \rm{a.u.}$. The Mathieu-type potential \cite{Slater} is a typical model potential, and has been used extensively in the optical lattice research area \cite{Breid,Chang} and recent solid HHG studies \cite{Wu2016,Wu2015,Guan,Du,Du-PRL}. We perform all calculations in the coordination space in the region [-240, 240] a.u. (60 lattice periods).

The energy band structure of a crystal is obtained by solving the eigenvalue equation of field-free Hamiltonian $\hat{H}_0$
\begin{eqnarray} \label{H0}
    \hat{H}_0\phi_n(x) = E_n\phi_n(x),
\end{eqnarray}
where $n$ is the eigenstate number and $\phi_n(x)$ is the corresponding eigenstate wavefunction. We numerically solve Eq. (\ref{H0}) by diagonalizing $\hat{H}_0$ on a coordinate grid. Specifically, the $\hat{H}_0$ operator is represented by the $N\times N$ square matrix $\mathbf{H}$, where $N$ is the number of grid points. The nonzero elements of the matrix $\mathbf{H}$ are given by
\begin{equation}\label{diagonalization}
\begin{split}
& \mathbf{H}_{i,i} = \frac{1}{(\Delta x)^2}+V_{i}, \\
& \mathbf{H}_{i,i+1} = -\frac{1}{2(\Delta x)^2}, \\
& \mathbf{H}_{i+1,i} = \mathbf{H}_{i,i+1}, \\
\end{split}
\end{equation}
where $\Delta x$ is the grid spacing and $V_{i}$ is the $i$-th element of the one-dimensional grid of $V$. The eigenenergy $E_n$ and eigenstate $\phi_n(x)$ are obtained by calculating the eigenvalues and eigenvector of matrix $\mathbf{H}$.

\begin{figure}[htb]
    \centerline{
        \includegraphics[width=9cm]{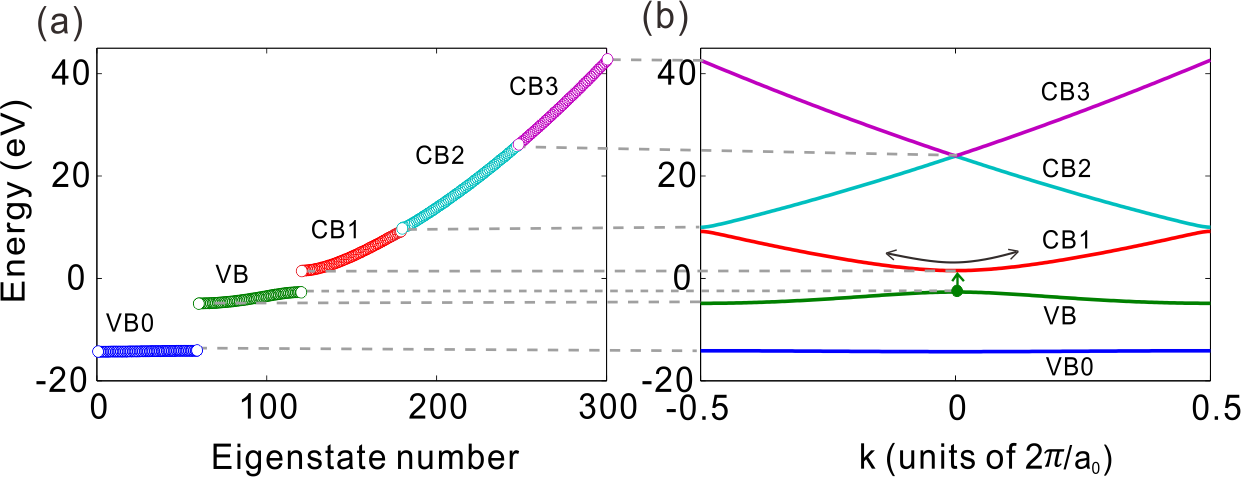}}
    \caption{The band structures calculated by (a) diagonalization scheme in coordinate space and by (b) Bloch states expansion in reciprocal space. Five bands are shown in the figures.}
    \label{fig1}
\end{figure}

Figure \ref{fig1}(a) shows the band structures calculated with the diagonalization scheme, where the band groups can be clearly distinguished. As illustrated in Fig. \ref{fig1}(a), the five bands are denoted as VB0, VB, CB1, CB2, and CB3, respectively.  The state numbers corresponding to the five bands are 1--59, 60--120, 61--180, 181--240 and 241--300, respectively. In order to verify the accuracy of resulting bands, we calculate the band structure by using the Bloch state basis \cite{Wu2015}. The obtained band structure is shown in Fig. \ref{fig1}(b). One can see that the features of bands (such as the number of bands and the energy range of each band) obtained with the two methods are in good agreement, which confirms the accuracy of resulting ground states and field-free bands.

When solids are irradiated by a laser pulse, electrons in the valence bands have opportunities to tunnel into conduction bands. The tunneling probabilities exponentially decay with the increase of bandgap. Considering the laser parameters used in current works, only a small proportion of electrons near $k=0$ in valence band VB can tunnel into conduction bands (as indicated by the green arrow in Fig. \ref{fig1}(b)). Therefore, we choose the eigenstate with $k=0$ in VB as the initially populated state. Since the lowest band VB0 is very flat and deeply bound, it plays negligible roles in the HHG dynamics. The time-dependent wavefunction $\psi(t)$ is obtained by solving the time-dependent Schr\"{o}dinger equation (TDSE) using split-operator technique \cite{Feit}. The time step is 0.03 a.u.. An absorbing boundary is adopted to overcome the unphysical reflections at the edges of the grid. In our calculation, the width of the absorbing boundary is adopted as 40 a.u.. The wavelengths of the driving laser pulses are restricted in MIR region. We adopt a sin$^2$ envelope for all laser pulses in this works.

The harmonic spectrum is obtained by Fourier transforming the laser-induced current:
\begin{eqnarray} \label{Hw}
H(\omega) = \left|\int{j(t)e^{i\omega t}dt}\right| ^2,
\end{eqnarray}
where the laser-induced current $j(t)$ is given by \cite{Guan}
\begin{eqnarray} \label{jt}
j(t) = -\left< \psi(t)\mid \hat{p} \mid  \psi(t)\right>.
\end{eqnarray}
In order to improve the signal-noise ratio, we multiply $j(t)$ by a Hanning window \cite{Wu2015} before the Fourier transform. The signal-noise ratio could be low because the laser intensities adopted in present work are relatively low.

To obtain the TDPI that reveals the HHG process in solids, instantaneous populations of the electrons on each eigenstate should be calculated during the TDSE propagation. The instantaneous population $\mid C_n(t) \mid ^2$ on eigenstate $\phi_n$ is obtained by calculating the modulus square of the time-dependent projection of $\psi(t)$ on $\phi_n$ as:
\begin{eqnarray} \label{cn}
\mid C_n(t) \mid ^2 \ = \ \mid \left< \phi_n \mid \psi(t)\right> \mid ^2.
\end{eqnarray}
Since $\phi_n$ corresponds to various eigenenergies $E_n$, $\mid C_n(t)\mid^2$ can also be understood as the time-dependent probability of electrons occupying on the energy level $E_n$. Then the TDPI is obtained by plotting $\mid C_n(t) \mid^2$  as a function of time $t$ and $E_n$. In TDPIs, electronic dynamics in HHG can be clearly observed from the energy-resolved population evolution of electrons.

\noindent \section{start and cutoff energies of plateaus} \label{TDPI}
\begin{figure}[htb]
    \centerline{
        \includegraphics[width=8cm]{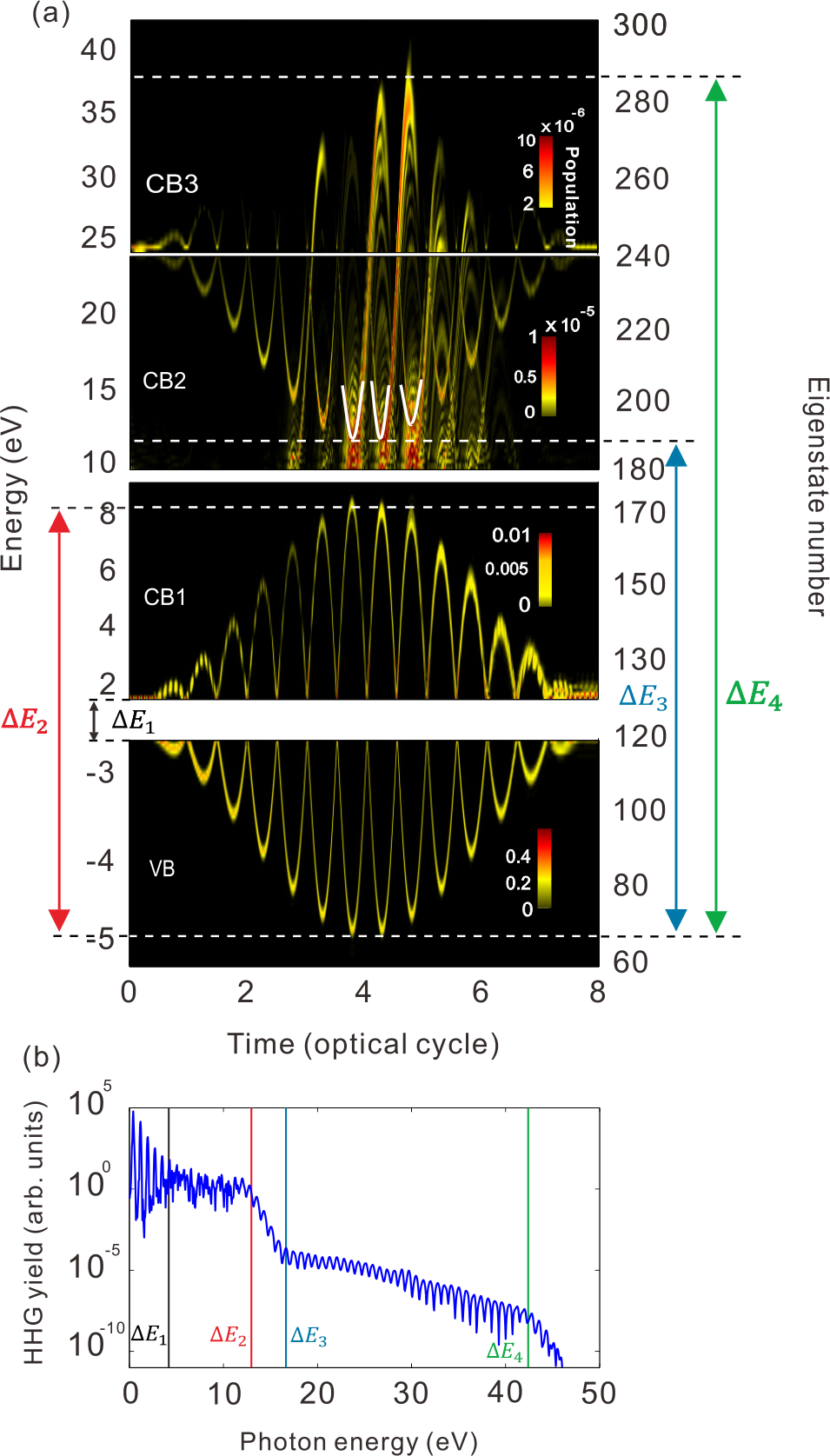}}
    \caption{The (a) TDPI and (b) harmonic spectra obtained with laser wavelength $\lambda=3.20\ \mu$m and the laser intensity $\rm I=8.09 \times 10^{11} W/cm^2$. The total pulse duration is 8 cycles. In the TDPI, the horizontal white dashed lines indicate the maximum or minimum instantaneous energies of the oscillating electrons. The oscillation peaks in CB2 are indicated by the white solid lines for much clear observations. The vertical solid lines in harmonic spectra indicate the instantaneous energy differences obtained from the TDPI.}
    \label{fig2}
\end{figure}

Figure \ref{fig2}(a) shows the calculated TDPI for a HHG process driven by the laser pulse with wavelength $\lambda=3.20$ $\mu$m and intensity $I=\rm 8.09 \times 10^{11}\ W/cm^2$. The total duration of the laser pulse adopted is 8 optical cycles. Several features can be found from the TDPI in Fig. \ref{fig2}(a). One can see obvious energy-resolved electron population oscillations in respective bands. These population oscillations correspond to the laser-driving Bloch oscillations of electrons in reciprocal space, where the electrons are driven forth and back periodically by the external laser field (as indicated by the black arrows in Fig. \ref{fig1}(b)). The strong oscillations shown in TDPIs also indicate that the solid HHG is a highly delocalized process, unlike the gas HHG where electrons are mainly localized in the ground state. The profiles of population oscillations are clear and bright in VB and CB1. However, the profiles become blurring in CB2 and CB3. This is because it is more difficult for laser-driven electrons to be populated into these higher conduction bands, and therefore the corresponding electron populations are about three orders of magnitude lower than that in CB1 as shown in Fig. 2(a). Due to the low electron populations, some evident strip-like structures appear at the bottom of CB2 when electrons in CB1 oscillate close to the top of the band. Considering that the signals of the population oscillations are blurred by the strip-like structures, we indicate the oscillation peaks using white solid curves in the blurred region.

The obtained harmonic spectrum is shown in Fig. \ref{fig2}(b). One can clearly see a characteristic two-plateau structure as described in Ref. \cite{Wu2015,Guan}, where each plateau has a start and a cutoff. The intensity of the second plateau is about five orders of magnitude lower than that of the first plateau. Similar to previous studies \cite{Wu2015,Vampa-PRL,Tamaya}, the harmonic spectrum exhibits clean odd harmonics in low energy region and noisy continuum-like structures in both first and second plateaus. The absence of clean harmonics in plateaus can be ascribed to several reasons, such as the infinitely long dephasing time \cite{Vampa-PRL}, elastic or inelastic scattering processes \cite{Kemper}, \emph{etc}. According to the studies by Wu {\it et al.} \cite{Wu2016,Wu2015}, the first plateau originates from the interband transitions from CB1 to VB, and the second plateau is due to the interband transitions from CB2 and CB3 to VB. Specifically, for the first plateau, the oscillating electron shown in the TDPI undergoes a transition from CB1 to VB, accompanied by the emission of a harmonic photon. Similarly, the second plateau is contributed by a transition of oscillating electrons from CB2 and CB3 to VB. In the discussions, the CB2 and CB3 are considered as a whole because CB2 and CB3 are very close and strongly coupled to each other. The real-time photon energy of the emitted harmonic is equal to the instantaneous energy difference of oscillating electrons in corresponding bands.

Based on the viewpoint mentioned above, the start and cutoff energies of both the first and second plateaus can be predicted exactly from TDPIs. In following discussions, the instantaneous energies corresponding to the population oscillation in VB, CB1, CB2 and CB3 shown in TDPIs are denoted as $\mathcal{E}_{\rm {VB}}(t)$, $\mathcal{E}_{\rm {CB1}}(t)$, $\mathcal{E}_{\rm {CB2}}(t)$ and $\mathcal{E}_{\rm {CB3}}(t)$, respectively. When electrons undergo the transitions from CB1 to VB, the minimum energy difference $\Delta E_1$ and maximum energy difference $\Delta E_2$ are obtained respectively by
\begin{align}
& \Delta E_1 = {\rm min}[\mathcal{E}_{\rm {CB1}}(t)-\mathcal{E}_{\rm {VB}}(t)], \label{D_E1} \\
& \Delta E_2 = {\rm max}[\mathcal{E}_{\rm {CB1}}(t)-\mathcal{E}_{\rm {VB}}(t)]. \label{D_E2}
\end{align}
As shown in Fig. \ref{fig2}(a), $\Delta E_1$ and $\Delta E_2$ can be easily read from TDPI. $\Delta E_1$ is equal to the bandgap between VB and CB1. $\Delta E_2$ is the energy difference between the highest (or deepest) peaks of population oscillations in VB and CB1 as indicated by the white dashed lines. Considering that the first plateau originates from the transitions from CB1 to VB, $\Delta E_1$ and $\Delta E_2$ should correspond to the start and cutoff energies of the first plateau, respectively. In Fig. \ref{fig2}(b), $\Delta E_1$ and $\Delta E_2$ are indicated by the black and red solid lines in harmonic spectrum, respectively. One can see that positions of $\Delta E_1$ and $\Delta E_2$ agree very well with the start and cutoff of the first plateau respectively. Similarly, for the transitions from CB2 and CB3 to VB, the minimum energy $\Delta E_3$ and maximum energy $\Delta E_4$ are given by
\begin{align}
& \Delta E_3 = {\rm min}[\mathcal{E}_{\rm {CB2}}(t)-\mathcal{E}_{\rm {VB}}(t)], \label{D_E3} \\
& \Delta E_4 = {\rm max}[\mathcal{E}_{\rm {CB3}}(t)-\mathcal{E}_{\rm {VB}}(t)]. \label{D_E4}
\end{align}
$\Delta E_3$ and $\Delta E_4$ can also be easily found as shown in Fig. \ref{fig2}(a). The second plateau is caused by the transitions from CB3 and CB2 to VB. Therefore, as shown in Fig. \ref{fig2}(b) by cyan and green solid lines, $\Delta E_3$ and $\Delta E_4$ match the start and cutoff energies of the second plateau accurately, respectively. The result shows that the start and cutoff energies of harmonic plateaus can be extracted accurately from the corresponding TDPIs. This conclusion has also been verified by our more simulations with different laser parameters. In addition, the low harmonic intensity of the second plateau can be interpreted by the low populations in CB2 and CB3. There are fewer electrons contributing to the transition back to VB in CB2 and CB3 than in CB1.

\begin{figure}[htb]
    \centerline{
        \includegraphics[width=8cm]{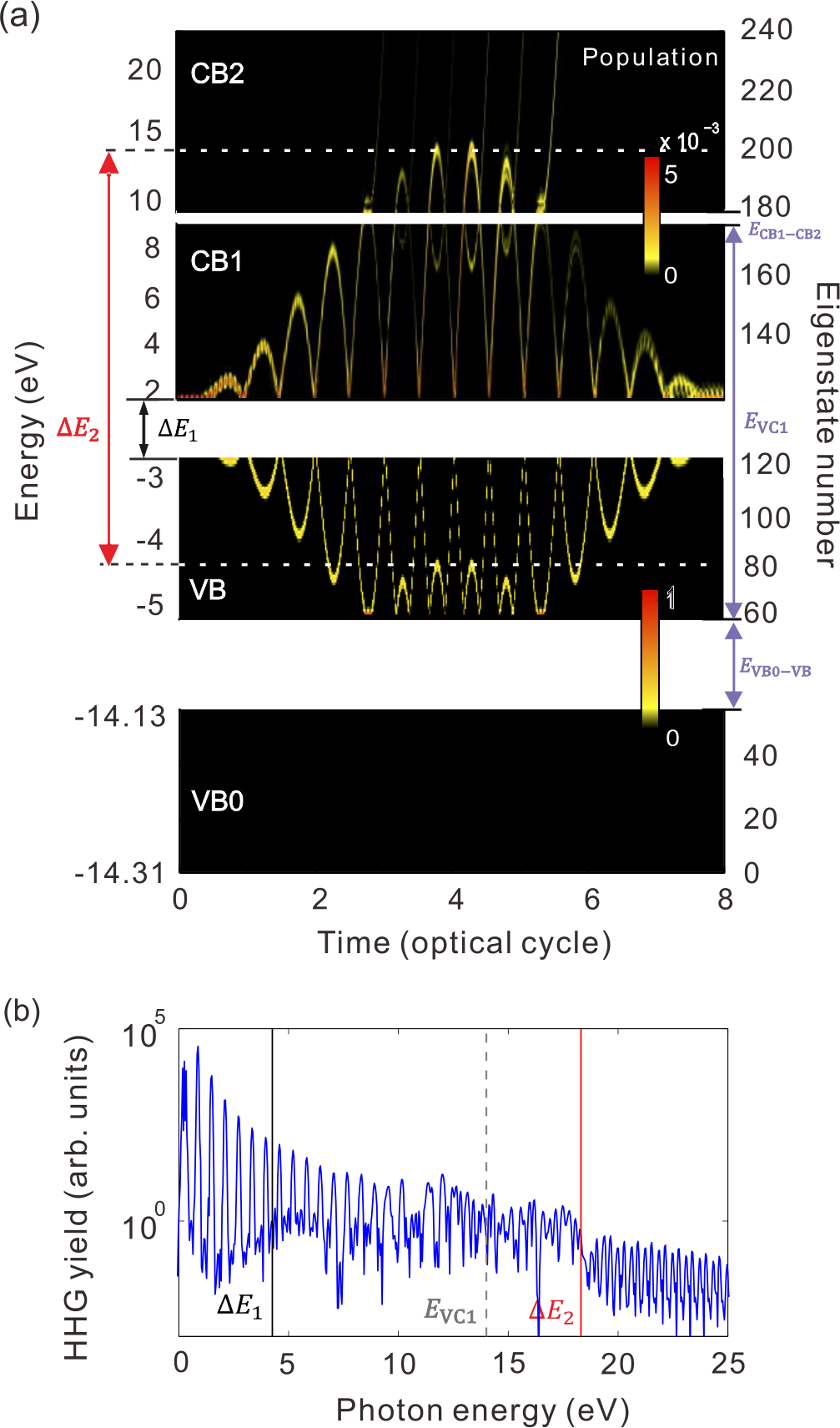}}
    \caption{The (a) TDPI and (b) harmonic spectrum obtained with laser wavelength $\lambda=4.00$ $\mu$m and laser intensity $ I=1.20 \rm \times 10^{12}\ W/cm^2$. The vertical solid lines in the harmonic spectrum indicate the instantaneous energy differences obtained from TDPIs. The gray dashed line indicate the total span of VB and CB1 $E_{\rm VC1}$.}
    \label{fig3}
\end{figure}

In previous discussions, the population oscillations of electrons are confined within respective bands. Then, a question arises: will the cutoff energy of a plateau be confined by the energy range of the involved bands (including bandgaps)? For example, we denote the total span between the bottom of VB and top of CB1 as $E_{\rm VC1}$ as indicated in Fig. \ref{fig3}(a) by a purple arrow. Will the cutoff energy of the first plateau not extend $E_{\rm VC1}$? In order to answer the question, we adopt the laser wavelength of $\lambda=4.00$ $\mu$m and laser intensity of $ I=1.20 \rm \times 10^{12}\ W/cm^2$ to calculate the TDPI and harmonic spectrum. Such longer wavelength and higher intensity can populate the electrons to higher levels. In following discussion, we will only focus on the first plateau since the second plateau is much weaker.

The obtained TDPI and harmonic spectrum are shown in Figs. \ref{fig3}(a) and \ref{fig3}(b), respectively. From Fig. \ref{fig3}(a), one can see that, when the oscillating electrons in CB1 reach the top of the band (the boundary of the first Brillouin zone in reciprocal space), most of the electrons can tunnel through the bandgap into CB2. The tunneling electrons in CB2 will continue oscillating driven by the laser field. In order to focus on the population oscillations of electrons tunneling from CB1, we adopt the same color scale in CB1 and CB2. The bandgap between CB1 and CB2 is very narrow ($E_{\rm CB1-CB2}$ = 0.84 eV). Therefore, the electrons in CB1 can easily tunnel to CB2. In contrast, as shown in Fig. \ref{fig3}(a), when oscillating electrons in VB reach the bottom of band (the boundary of the first Brillouin zone in reciprocal space), the electrons can hardly tunnel into VB0 since the bandgap $E_{\rm VB0-VB}$ is quite broad ($E_{\rm VB0-VB}$ = 9.25 eV). From the TDPI, one can see that, since the oscillating electrons can tunnel into higher bands with high probability, the possible maximum cutoff energy is not limited by the energy range of the bands. As shown in Fig. \ref{fig3}(a), the maximum energy difference $\Delta E_2$ for the first plateau is indicated by the red arrow, which spans from the highest oscillation peak in CB2 to that in VB. From Fig. \ref{fig3}(b), one can see that $\Delta E_2$ is in good agreement with the cutoff energy of the first plateau as shown in the harmonic spectrum. The gray dashed line shown in Fig. \ref{fig3}(b) indicates the position of $E_{\rm VC1}$ ($E_{\rm VC1}$ = 14.00 eV in our model). One can see that the cutoff energy (see $\Delta E_2$) is larger than $E_{\rm VC1}$, i.e., the possible maximum cutoff energy of the first plateau is larger than $E_{\rm VC1}$.

\noindent \section{emission time and trajectory analysis}\label{trajectories}
\begin{figure}[htb]
    \centerline{
        \includegraphics[width=8cm]{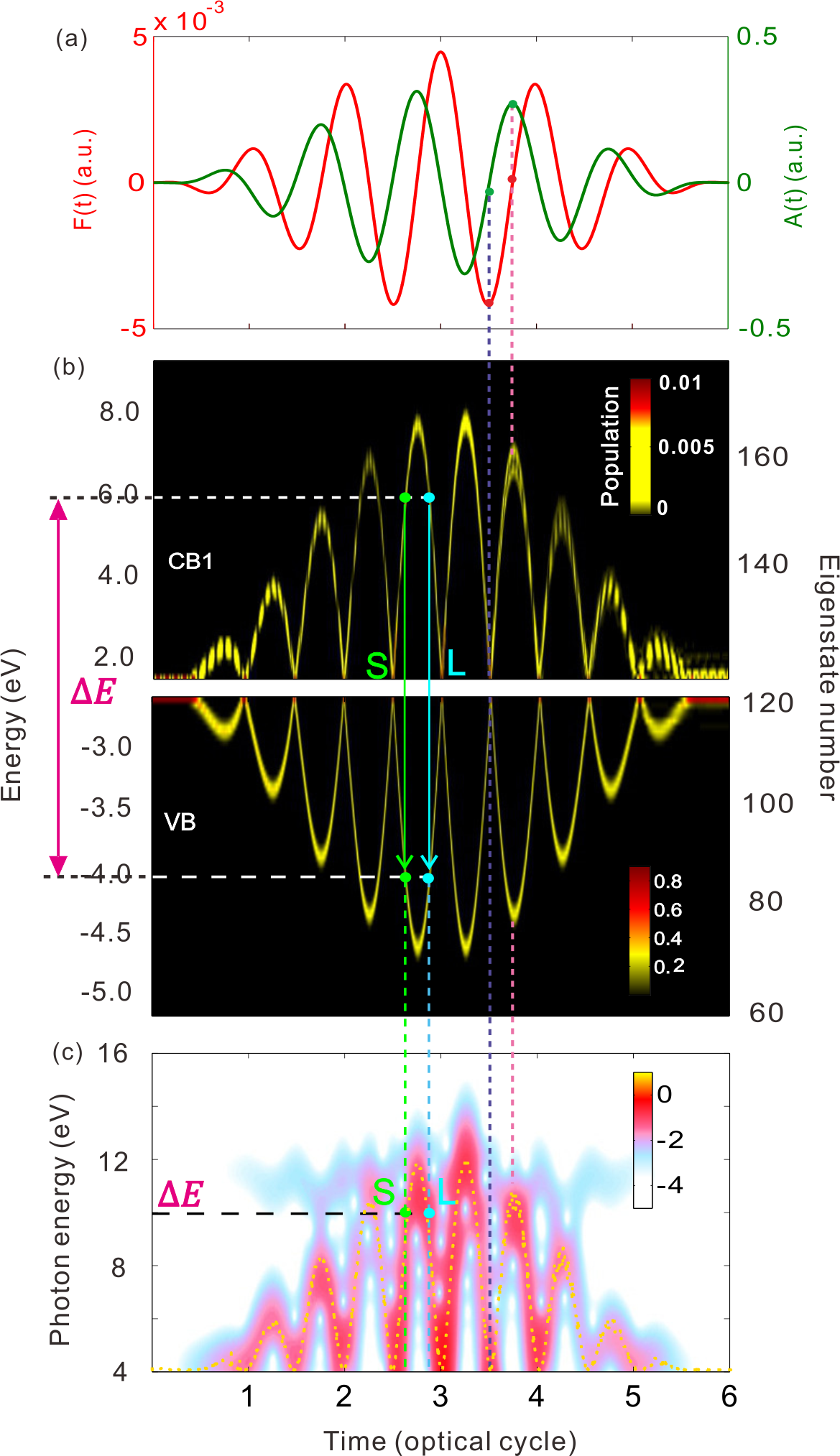}}
    \caption{(a) The electric field and vector potential of the laser pulse with laser wavelength $\lambda=3.20$ $\mu$m and laser intensity $ I=7.00 \rm \times 10^{11}\ W/cm^2$. The total duration is 6 optical cycles. (b) The TDPI obtained with the field shown in panel (a). (c) The time-frequency spectrum with logarithmic color scaling. The yellow dashed curve is the instantaneous energy difference of oscillating electrons in CB1 and VB obtained from the TDPI. In panels (b) and (c), S and L denote the short and long trajectories.}
    \label{fig4}
\end{figure}

In this section, the emission times of high harmonics in solids will be discussed with TDPIs. Figure \ref{fig4}(b) shows the calculated TDPI for a HHG process driven by a 3.20 $\mu$m laser pulse with intensity of $7.00 \rm \times 10^{11}\ W/cm^2$. The pulse duration is 6 optical cycles. The electric field and vector potential of the laser pulse are plotted in Fig. \ref{fig4}(a). In order to obtain the emission times of harmonics, we calculate the time--frequency (TF) spectrum by Gabor transform \cite{Chirila} of the time-dependent current. The resulting TF spectrum is shown in Fig. \ref{fig4}(c). One can see that the harmonic emissions occur four times per optical cycle. For each half cycle, the emissions correspond to a pair of short and long branches (labeled as S and L respectively). From Figs. \ref{fig4}(a)-\ref{fig4}(c), it is found that the evolution of the time-dependent population in TDPI agrees well with that of the TF spectrum. The peaks of population oscillations in the TDPI and HHG radiations in the TF spectrum both correspond to the zero points of laser fields and peaks of vector potentials as indicated by the pink dashed lines. On the contrary, the minima of population oscillations and HHG radiations both correspond to the peaks of laser fields and zero points of vector potentials as indicated by purple dashed lines. Furthermore, we calculate the instantaneous energy differences of oscillating electrons in CB1 and VB, i.e., $\mathcal{E}_{\rm {C1-V}}(t)$. $\mathcal{E}_{\rm {C1-V}}(t)$ is obtained from the TDPI shown in Fig .\ref{fig4}(b) as
\begin{eqnarray} \label{C1-V}
\mathcal{E}_{\rm {C1-V}}(t) = \mathcal{E}_{\rm {CB1}}(t)-\mathcal{E}_{\rm {VB}}(t).
\end{eqnarray}
In Fig. \ref{fig4}(c), we plot $\mathcal{E}_{\rm {C1-V}}(t)$  as the yellow dashed curve. One can see that $\mathcal{E}_{\rm {C1-V}}(t)$ is consistent with the harmonic signals shown in TF spectrum very well. Specifically, both of the short and long branches shown in TF spectrum coincide with the $\mathcal{E}_{\rm {C1-V}}(t)$ curve. In the studies for gas HHG, the TF spectrum is the one of the most important and frequently used tools to analyze the HHG dynamics. The correspondence between the TDPI and TF spectrum indicates that the TDPI is a powerful tool to analyze the HHG processes in solids.

In gas HHG, semi-classical three-step model builds up an intuitive picture to describe the HHG process \cite{Corkum_2}. In this picture, the short and long trajectories are distinguished according to the duration for the tunneling electrons traveling in the continuum \cite{Lewenstein}. In solid HHG, as shown in Fig. \ref{fig4}(c), the ``short'' and ``long'' trajectories are still visible and correspond to the short and long branches in TF spectrum, respectively. From the TDPI, it is shown that the concepts of short and long trajectories for solid HHG can be established according to the short and long branches of the population oscillations within the same half cycle. For example, as shown in Fig. \ref{fig4}(b), for the harmonic with photon energy $\Delta E = 9.96$ eV, the emission occurs twice per half cycle. As donated by the green arrow, the transition from the short branch occurs at time $t_S = 2.63 \ T_0$, where $T_0$ represents one optical cycle. Then the electrons oscillate to higher energies and return back to the same energy level at $t_L = 2.87 \ T_0$, and harmonic photon with the same energy $\Delta E = 9.96$ eV is emitted via the transition from the long branch as donated by a cyan arrow. The emission time $t_S$ is earlier than $t_L$. Therefore, the emission pathways indicated by green and cyan arrows can be called short and long trajectories, respectively. The short and long trajectories read from TDPIs are consistent with those from TF spectrum. As shown in Fig. \ref{fig4}(c), $t_S$ and $t_L$ are in good agreement with the emission times for $\Delta E = 9.96$ eV read from the short and long branches of TF spectrum, respectively. Furthermore, similar to the gas HHG, the short trajectory in solid HHG is positively chirped, whereas the long trajectory is negatively chirped as shown in Fig. \ref{fig4}(b) and \ref{fig4}(c). In the reciprocal space, the short and long trajectories appear essentially because the electrons are driven forth and back during the Bloch oscillation and will pass the same point (with specific $k$) twice in one half cycle.

\begin{figure}[htb]
    \centerline{
        \includegraphics[width=8cm]{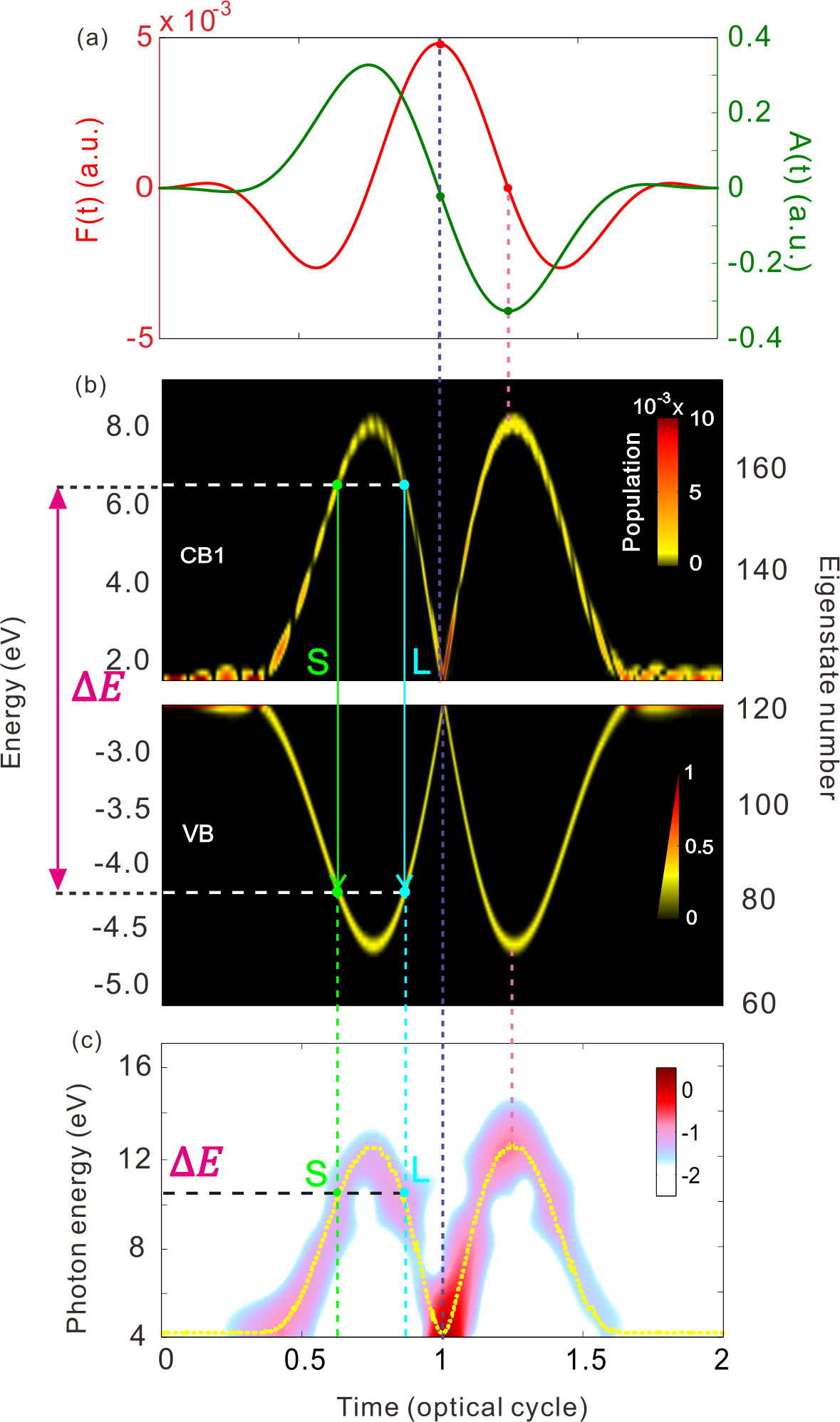}}
    \caption{(a) The electric field and vector potential of the laser pulse with laser wavelength $\lambda=3.20$ $\mu$m and laser intensity $ I=8.09 \rm \times 10^{11}\ W/cm^2$. The total duration is 2 optical cycles. (b) The TDPI obtained with the field shown in panel (a). (c) The time-frequency spectrum with logarithmic color scaling. The yellow dashed curve is obtained by the instantaneous energy difference of oscillating electrons in CB1 and VB. In panels (b) and (c), S and L denote the short and long trajectories.}
    \label{fig5}
\end{figure}

We also adopt a few-cycle pulse to demonstrate the short and long trajectories of HHG in solids. With the few-cycle pulse, the short and long trajectories can be more distinguishable. The applied wavelength is $\lambda=3.20$ $\mu$m and the intensity is $ I=8.09 \rm \times 10^{11}\ W/cm^2$. The total duration of the laser pulse is 2 optical cycles. The electric field and vector potential are shown in Fig. \ref{fig5}(a). Figures \ref{fig5}(b) and \ref{fig5}(c) show the obtained TDPI and TF spectrum, respectively. As in the previous discussion, one can see that $\mathcal{E}_{\rm {C1-V}}(t)$ agrees with the TF spectrum in Fig. \ref{fig5}(c) very well. The short and long trajectories are recognized more clearly in Fig. \ref{fig5}(b). Here, we take the harmonic with photon energy $\Delta E = 10.74 $ eV as an example. The corresponding emission times $t_S$ and $t_L$ obtained form the TDPI agree very well with those read from the TF spectrum. The results and discussions in this section shows that the TDPI provides an intuitive picture to describe real-time HHG dynamics in solids. The information of harmonic radiations can be directly extracted from the corresponding TDPI.

\noindent \section{Carrier--envelope phase effect} \label{CEP}
For a few-cycle laser pulse, the CEP will dramatically affect the temporal shape of the electric field. Some physical processes induced by the few-cycle laser field will rely on the variation of CEP. For example, the cutoff energy of the gas HHG sensitively depends on the CEP of the driving field \cite{Nisoli,Bohan}. The CEP effect have been discussed widely for photoionization \cite{Dietrich,Christov} and gas HHG \cite{Baltuska,Carrera}.

\begin{figure}[htb]
    \centerline{
        \includegraphics[width=8.5cm]{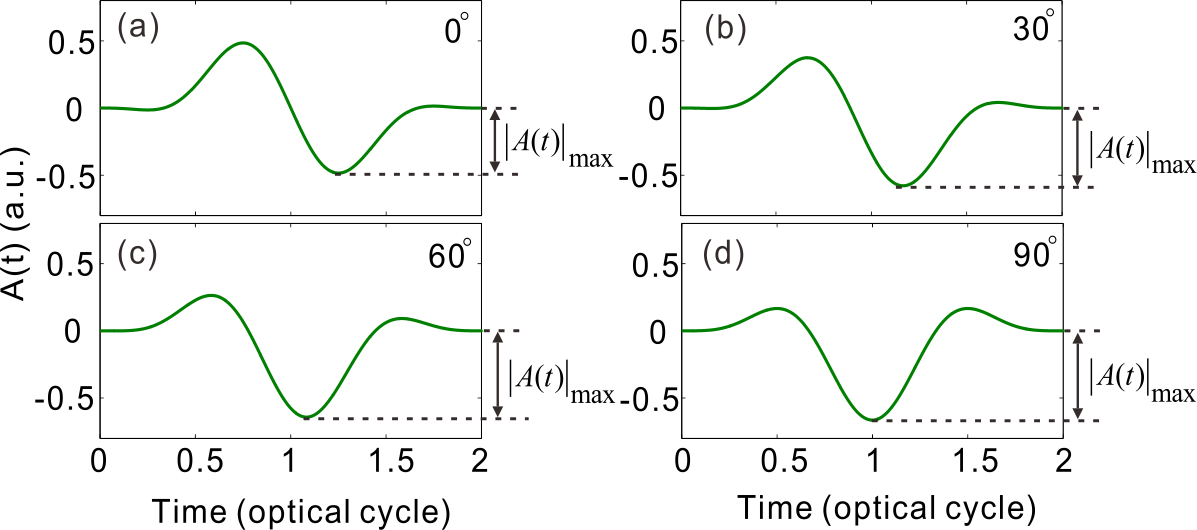}}
    \caption{Vector potentials of laser pulses with different CEPs (a) $\phi=0^{\circ}$. (b) $\phi=30^{\circ}$. (c) $\phi=60^{\circ}$. (d) $\phi=90^{\circ}$. The laser wavelength is $\lambda=3.60$ $\mu$m and the laser intensity is $ I=1.40 \rm \times 10^{12}\ W/cm^2$. The total duration is 2 optical cycles. $|A(t)|_{\max}$ represents the maximum value of module of vector potentials $A(t)$.}
    \label{fig6}
\end{figure}

\begin{figure}[htb]
    \centerline{
        \includegraphics[width=7cm]{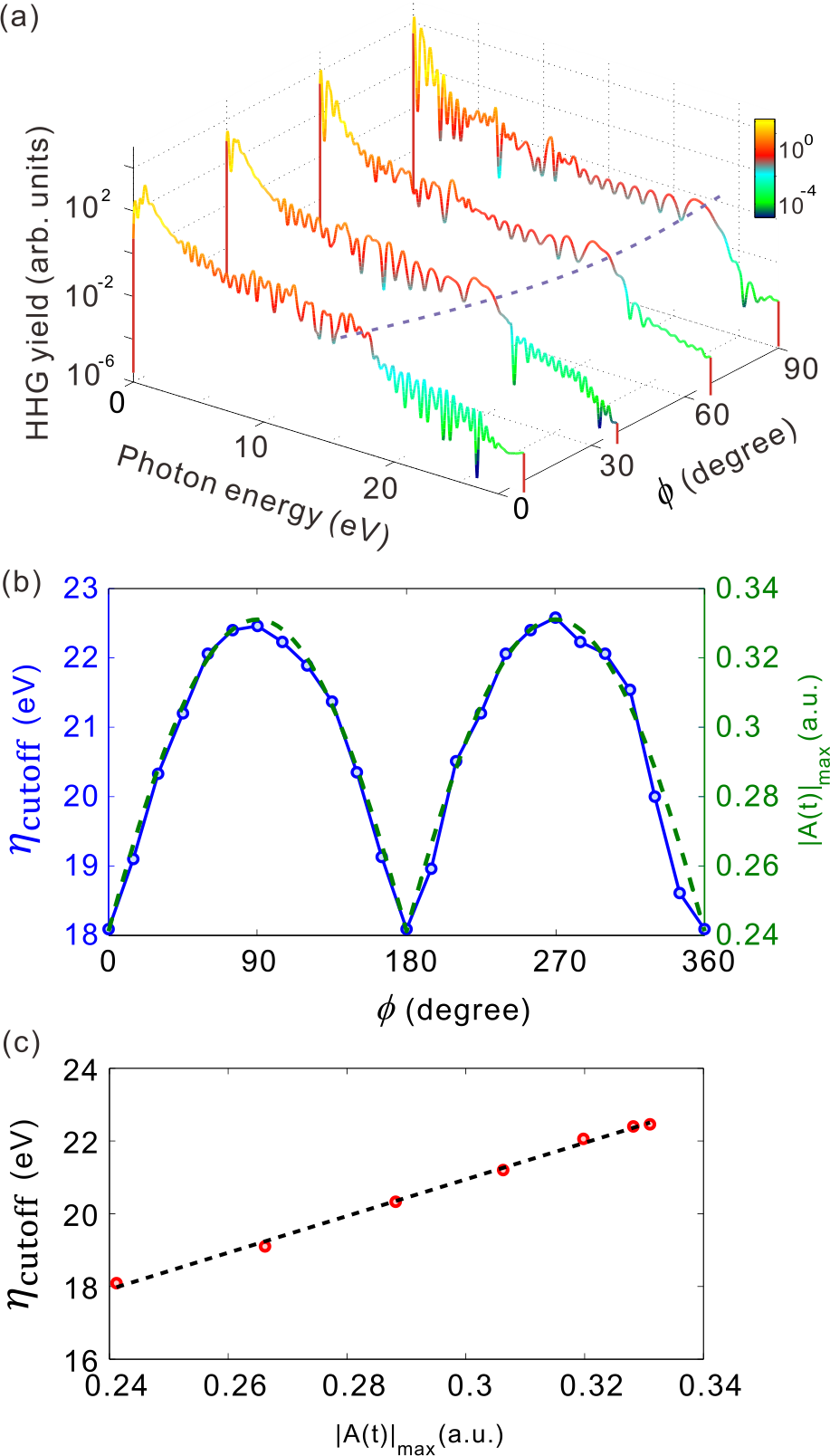}}
    \caption{(a) High harmonic spectra with $\phi=0^{\circ}, 30^{\circ}, 60^{\circ}$ and $90^{\circ}$. The purple dashed line indicates the cutoff regions. (b) The blue solid line shows the CEP dependence of the cutoff energy $\eta_{\rm cutoff}$. The green dashed line shows the $|A(t)|_{\max}$. (c) $\eta_{\rm cutoff}$ as a function of $|A(t)|_{\max}$, where $\phi$ is chosen from $0^{\circ}$ to $90^{\circ}$.}
    \label{fig7}
\end{figure}

In this section, we will analyze the CEP effect for solid HHG by using the TDPI. The adopted laser wavelength is $\lambda=3.60$ $\mu$ and laser intensity is $ I=1.40 \rm \times 10^{12}\ W/cm^2$. The total duration of the laser pulse is 2 optical cycles. Figures \ref{fig6}(a)-\ref{fig6}(d) show the vector potentials of the laser pulses with CEP $\phi=0^{\circ}, 30^{\circ}, 60^{\circ}$ and $90^{\circ}$ respectively. The maximum of module of vector potentials $|A(t)|_{\max}$ are indicated with the vertical arrows. In the following discussion, we focus on the first cutoff and the cutoff energy is denoted as $\eta_{\rm cutoff}$.

\begin{figure*}[htb]
    \centerline{
        \includegraphics[width=16cm]{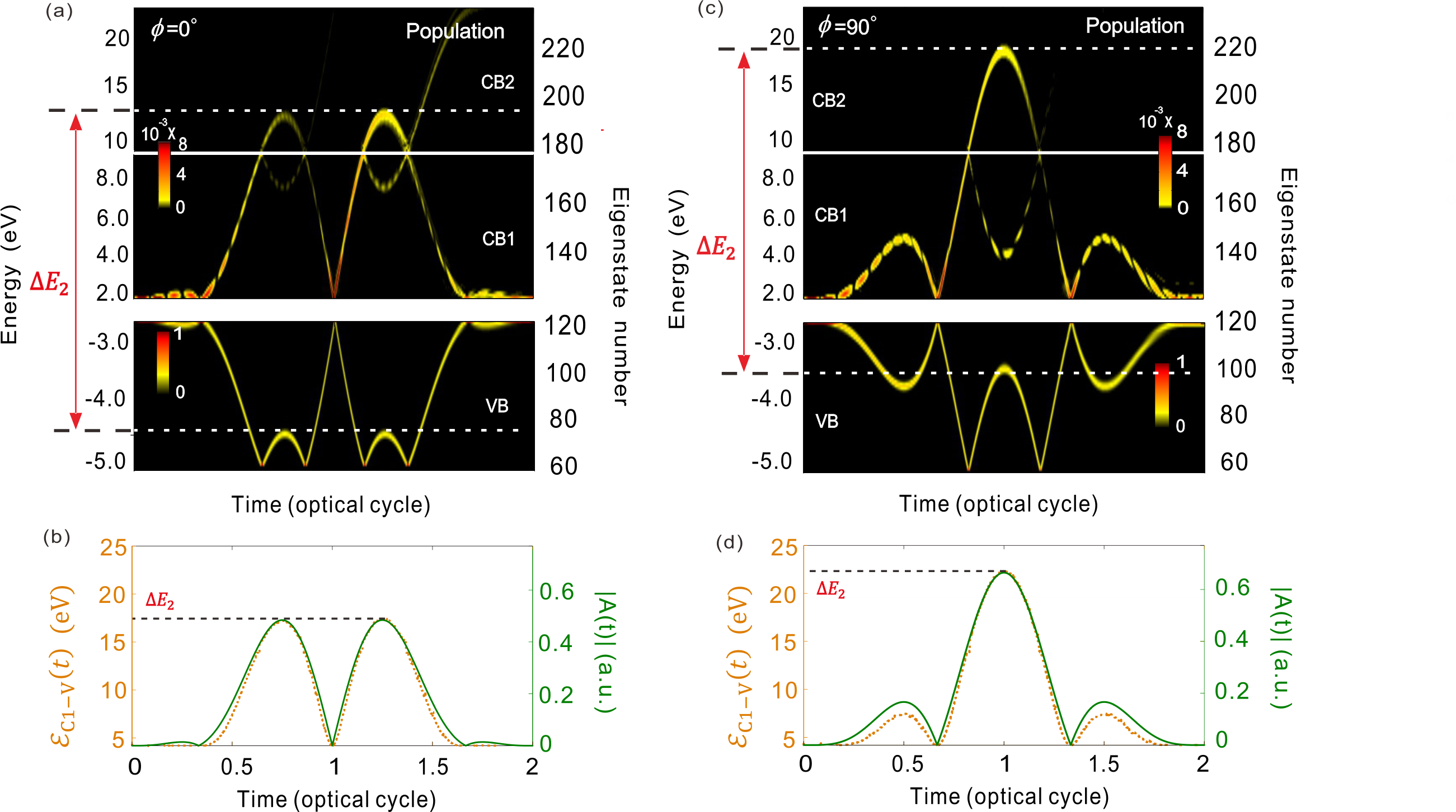}}
    \caption{(a) The TDPI obtained with $\phi = 0^{\circ}$. (b) The real-time radiation energy $\mathcal{E}_{\rm {C1-V}}(t)$ and $|A(t)|$ for $\phi = 0^{\circ}$. (c) The TDPI obtained with $\phi = 90^{\circ}$. (d) The real-time radiation energies $\mathcal{E}_{\rm {C1-V}}(t)$ and $|A(t)|$ for $\phi = 90^{\circ}$. In panels (b) and (d), the yellow dashed lines represent $\mathcal{E}_{\rm {C1-V}}(t)$, and the green solid lines represent $|A(t)|$.}
    \label{fig8}
\end{figure*}

High harmonic spectra for $\phi=0^{\circ}, 30^{\circ}, 60^{\circ}$ and $90^{\circ}$ are shown in Fig. \ref{fig7}(a). One can see clearly that the cutoff energy $\eta_{\rm cutoff}$ is very sensitive to the CEP. Specifically, as indicated by the purple dashed line, $\eta_{\rm cutoff}$ increases monotonously when $\phi$ varies from $0^{\circ}$ to $90^{\circ}$. In order to further discuss the relation of $\eta_{\rm cutoff}$ and $\phi$, we calculate the cutoff energy $\eta_{\rm cutoff}$ as a function of $\phi$ from $0^{\circ}$ to $360^{\circ}$ in step of $15^{\circ}$. The obtained result is shown in the Fig. \ref{fig7}(b) with the blue solid curve. Here, $\eta_{\rm cutoff}$ is read from the $\Delta E_2$ shown in the corresponding TDPI, since the $\eta_{\rm cutoff}$ is equal to maximum energy difference $\Delta E_2$ as discussed in Sec. \ref{TDPI}. One can see that $\eta_{\rm cutoff}$ exhibits a tendency of monotonous increasing in the region of $0^{\circ}$--$90^{\circ}$, which corresponds to the results shown in Fig. \ref{fig7}(a). Further studies show that $\eta_{\rm cutoff}$ has a close relation with $|A(t)|_{\max}$. In Fig. \ref{fig7}(b), the green dashed curve shows the $|A(t)|_{\max}$ as a function of $\phi$. One can see that $|A(t)|_{\max}$ and $\eta_{\rm cutoff}$ curves are nearly in complete agreement with each other. This indicates that the $\eta_{\rm cutoff}$ is determined by the $|A(t)|_{\max}$. Considering that the $\eta_{\rm cutoff}$ and $|A(t)|_{\max}$ curves are plotted with different linear vertical axes, it can be speculated that $\eta_{\rm cutoff}$ depends linearly on $|A(t)|_{\max}$, i.e.,
\begin{eqnarray} \label{cutof--Amax}
\eta_{\rm cutoff} \propto |A(t)|_{\max}.
\end{eqnarray}
Figure \ref{fig7}(c) shows the $\eta_{\rm cutoff}$ as a function of $|A(t)|_{\max}$, where $\phi$ is chosen in the range of $0^{\circ}$--$90^{\circ}$. The result confirms that $\eta_{\rm cutoff}$ increases linearly with $|A(t)|_{\max}$. For a long pulse, since the CEP only slightly influence the waveform of vector potential, $|A(t)|_{\max} = A_0$ for any value of $\phi$, where $A_0$ is the amplitude of vector potential. Thus, the cutoff energy for long pulses satisfies $\eta_{\rm cutoff} \propto A_0$ as discussed in Ref. \cite{Wu2016,Wu2015,Guan,Du-PRL}.

In the following, we choose $\phi=0^{\circ}$ and $\phi=90^{\circ}$ as examples to analyze the CEP effect using TDPIs. The TDPIs for $\phi=0^{\circ}$ and $\phi=90^{\circ}$ are shown in Figs. \ref{fig8}(a) and \ref{fig8}(c), respectively. The corresponding $\mathcal{E}_{\rm {C1-V}}(t)$ and $|A(t)|$ for $\phi=0^{\circ}$ and $\phi=90^{\circ}$ are shown in Figs. \ref{fig8}(b) and \ref{fig8}(d), respectively. From Figs. \ref{fig8}(a)-\ref{fig8}(d), one can see that the population oscillations for $\phi=0^{\circ}$ and $\phi=90^{\circ}$ are significantly different, but both of them are determined by respective $|A(t)|$. Specifically, as shown in Fig. \ref{fig8}(a), the population oscillation in CB1 exhibits two peaks with equal height. This trend is similar to that of the corresponding $|A(t)|$ curve shown in Fig. \ref{fig8}(b). Likewise, as shown in Fig. \ref{fig8}(c), the population oscillation in CB1 exists a prominent peak in the middle and two secondary peaks on both sides, which is similar to the trend of the $|A(t)|$ curve shown in Fig. \ref{fig8}(d). The correspondence of electronic oscillation and $|A(t)|$ essentially originates from the fact that the wave vector of Bloch electron depends linearly on the vector potential of the external laser field \cite{Du-PRL}.

Moreover, the module of vector potential not only dominates the population oscillations of electrons in their respective bands, but also governs the energy differences between conduction and valence bands, i.e., the real-time photon energies of emitted harmonics. As shown in Figs. \ref{fig8}(b) and \ref{fig8}(d), one can see that the $\mathcal{E}_{\rm {C1-V}}(t)$ has the same trend as $|A(t)|$ for both $\phi=0^{\circ}$ and $\phi=90^{\circ}$. Especially for the high-energy region, $\mathcal{E}_{\rm {C1-V}}(t)$ and $|A(t)|$ nearly coincide completely with each other. Considering different linear vertical axes are used for the two curves, it can be obtained for high-energy region that
\begin{eqnarray} \label{co--A}
\mathcal{E}_{\rm {C1-V}}(t) \propto |A(t)|.
\end{eqnarray}
Eq. (\ref{co--A}) is not only valid for the short pulse used here. It has been verified by our numerous other simulations. From Eq. (\ref{co--A}), Eq. (\ref{cutof--Amax}) can be obtained considering that $\eta_{\rm cutoff} = {\max}[\mathcal{E}_{\rm {C1-V}}(t)]$. Accordingly, since $|A(t)|_{\max}$ for $\phi=90^{\circ}$ is larger than that for $\phi=0^{\circ}$ as shown in Figs. \ref{fig8}(b) and \ref{fig8}(d), the cutoff energy for $\phi=90^{\circ}$ is larger than that for $\phi=0^{\circ}$.

Based on the above discussions, the CEP effect can be essentially understood in this picture: When the CEP varies, the corresponding variation of $A(t)$ leads to the variation of the Bloch oscillation in each bands. As a result, the maximum energy difference between the conduction band and the valence band varies and the cutoff changes. Specifically, the instantaneous photon energy of emitted harmonics in the high-energy region is proportional to $|A(t)|$, and the cutoff energy $\eta_{\rm cutoff}$ is proportional to $|A(t)|_{\max}$. As this picture can be intuitively revealed by the TDPI, the TDPI is a very useful tool to analyze the CEP effects in solid HHG.

\noindent \section{Two--color laser fields} \label{twocolor}

For the gas HHG, a lot of works have been devoted to studying the HHG in two--color laser fields \cite{Frolov,HHG_ll}, because the two--color field offers a powerful tool to regulate the HHG. For instance, the two-color field can be used to amplify the HHG yield and extend the harmonic cutoff \cite{Watanabe,LiuTT,ZhaiZ}. By varying the relative phase between the two components, the two-color field allows one to manipulate the HHG processes \cite{Ganeev,Ishii} and control the birth of attosecond XUV pulses \cite{Dudovich}. The dynamical processes of HHG in two-color fields are more complicated than in monochromatic field. To our knowledge, the HHG driven by two-color fields in solid phase was rarely investigated at present.

\begin{figure}[htb]
    \centerline{
        \includegraphics[width=8cm]{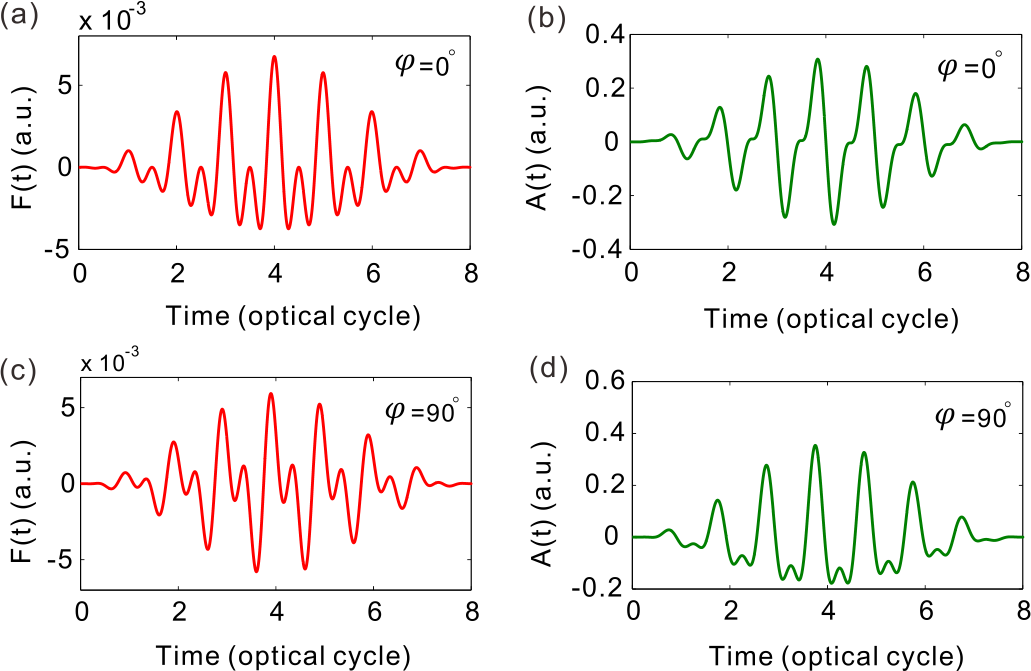}}
    \caption{The synthesized electric fields and vector potentials of the two-color laser pulse with (a) and (b) $\varphi = 0^{\circ}$, (c) and (d) $\varphi = 90^{\circ}$. The laser wavelengths of the fundamental and second harmonic fields are $\lambda_1=3.20$ $\mu$m and $\lambda_2=1.60$ $\mu$m, respectively. The laser intensities of the fundamental and second harmonic fields both are $ I=4.00 \rm \times 10^{11}\ W/cm^2$. The total duration of the laser pulse is 8 optical cycles.}
    \label{fig9}
\end{figure}

\begin{figure*}[htb]
    \centerline{
        \includegraphics[width=16cm]{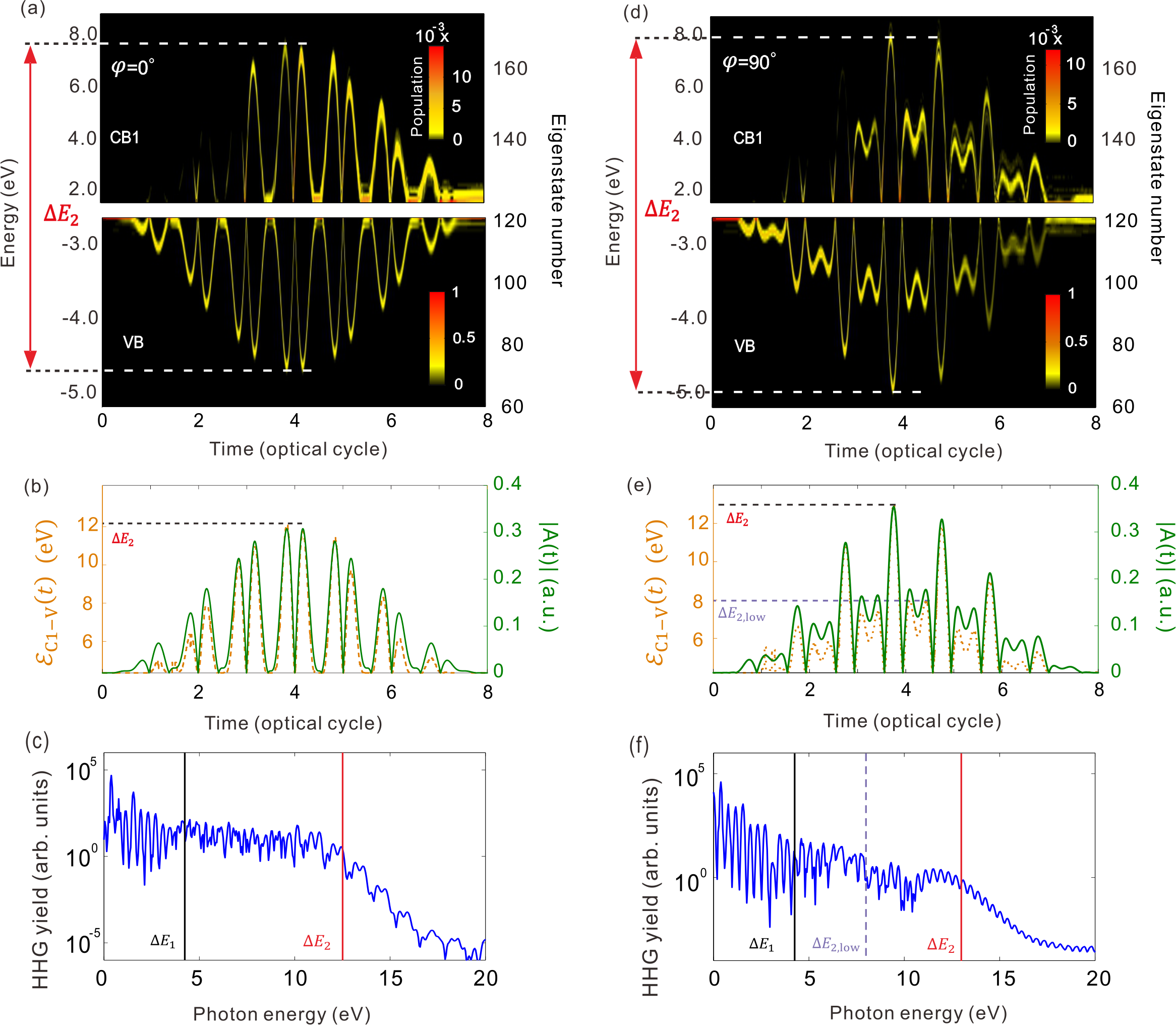}}
    \caption{(a) and (d) The TDPI obtained with $\varphi = 0^{\circ}$ and $90^{\circ}$, respectively. (b) and (e) Comparisons between the $\mathcal{E}_{\rm {C1-V}}(t)$ and $|A(t)|$ for $\varphi = 0^{\circ}$ and $90^{\circ}$, respectively. The yellow dashed curves represent the $\mathcal{E}_{\rm {C1-V}}(t)$ and the green solid curves represent the $|A(t)|$. (c) and (f) Harmonic spectra obtained with $\varphi = 0^{\circ}$ and $90^{\circ}$, respectively.}
    \label{fig10}
\end{figure*}

Herein, we demonstrate the HHG dynamics in solids driven by a two-color field involving a fundamental and a second harmonic field using TDPIs. In our calculation, the wavelengths of the fundamental and second harmonic fields are $\lambda_1=3.20$ $\mu$m and $\lambda_2=1.60$ $\mu$m, respectively. The laser intensities of two components  are both $ I=4.00 \rm \times 10^{11}\ W/cm^2$. The total duration of the laser pulse is 8 optical cycles of the fundamental field. The relative phase between the two fields is denoted as $\varphi$. Our analyses will focus on the situations of $\varphi = 0^{\circ}$ and $\varphi = 90^{\circ}$. The synthesized electric fields and vector potentials with $\varphi = 0^{\circ}$ and $90^{\circ}$ are shown in Figs. \ref{fig9}(a)-\ref{fig9}(d).

Figure \ref{fig10}(a) shows the TDPI for $\varphi=0^{\circ}$. Figure \ref{fig10}(b) shows the corresponding $\mathcal{E}_{\rm {C1-V}}(t)$ and $|A(t)|$. Figures \ref{fig10}(d) shows the TDPI for $\varphi=90^{\circ}$. Figure \ref{fig10}(e) shows the corresponding $\mathcal{E}_{\rm {C1-V}}(t)$ and $|A(t)|$. One can see that the population oscillations of electrons in respective bands are still clear as shown in Figs. \ref{fig10}(a) and \ref{fig10}(d). According to the discussions in Sec. \ref{CEP}, the photon energy of the emitted harmonics in the high-energy region depends linearly on $|A(t)|$. Hence, as shown in Figs. \ref{fig10}(b) and \ref{fig10}(e), $\mathcal{E}_{\rm {C1-V}}(t)$ curves are in good consistence with $|A(t)|$ curves for both $\varphi = 0^{\circ}$ and $\varphi = 90^{\circ}$. Compared with the population oscillations from the monochromatic field shown in Fig. \ref{fig2}(a), the population oscillations in Figs. \ref{fig10}(a) and \ref{fig10}(d) are peculiar. The special profiles of population oscillations are determined by the corresponding $|A(t)|$ as shown in Figs. \ref{fig10}(b) and \ref{fig10}(e).

The high harmonic spectra for $\varphi=0^{\circ}$ and $90^{\circ}$ are shown in Figs. \ref{fig10}(c) and \ref{fig10}(f), respectively. It is shown that the starts and cutoffs of the harmonic plateau are still in accord with $\Delta E_1$ and $\Delta E_2$ very well for both $\varphi=0^{\circ}$ and $90^{\circ}$. The cutoff energy for $\varphi=90^{\circ}$ is larger than that for $\varphi=0^{\circ}$, because $\Delta E_2$ for $\varphi=90^{\circ}$ is greater than that for $\varphi=0^{\circ}$ as shown in Figs. \ref{fig10}(a) and \ref{fig10}(d) (or Figs. \ref{fig10}(b) and \ref{fig10}(e)). This sensitivity of cutoff energies on relative phase $\varphi$ is essentially due to the change of $|A(t)|_{\max}$ when $\varphi$ varies.

In addition, the harmonic plateau for $\varphi=0^{\circ}$ is quite flat as shown in Fig. \ref{fig10}(c). This is because the profile of the population oscillation in the TDPI is composed of regular peaks similar to those for a monochromatic field.  By contrast, the plateau for $\varphi=90^{\circ}$ is relatively uneven as shown in Fig. \ref{fig10}(f). This uneven plateau is caused by the special structure of the profile of population oscillation in the TDPI and can be interpreted with the $\mathcal{E}_{\rm {C1-V}}(t)$. As shown in Fig. \ref{fig10}(e), the $\mathcal{E}_{\rm {C1-V}}(t)$ curve is composed of two kinds of peaks: the high sharp peaks and the low peaks with concave tops. Then,  $\mathcal{E}_{\rm {C1-V}}(t)$ curve can be divided into two parts by the maximum energy of the low peaks $\Delta E_{2,\rm low}$ as indicated by the purple dashed line in Fig. \ref{fig10}(e). The high harmonics with photon energy larger than $\Delta E_{2,\rm low}$ are emitted at most twice per cycle, whereas the high harmonics with photon energy smaller than $\Delta E_{2,\rm low}$ can be emitted four times per cycle. Therefore, HHG in the region below $\Delta E_{2,\rm low}$ is more efficient. In Fig. \ref{fig10}(f), $\Delta E_{2,\rm low}$ is indicated by the vertical purple dashed line. One can see that the intensity of harmonics ranging from $\Delta E_{2,\rm low}$ to $\Delta E_{2}$ is lower than that ranging from $\Delta E_{1}$ to $\Delta E_{2,\rm low}$.

The above results indicate that the two-color field can be also used to control the electronic dynamics in solids HHG as in gas HHG. The features of the generated high harmonics can be effectively modulated by adjusting the relative phase of two components. These modulation effects can be revealed clearly with TDPIs. Therefore, the TDPI provides a powerful tool to analyze the real-time HHG dynamics in solids driven by two-color fields and to guide people to manipulate the solid HHG.

\noindent \section{Conclusion} \label{conclusion}
In summary, this work introduces a intuitive representation called TDPI to reveal the real-time dynamics of solid HHG in a quantitative way. The population oscillations of electrons in their respective bands are intuitively demonstrated in TDPIs. We show that the real-time photon energies of harmonic radiations can be obtained directly from the instantaneous energy differences of oscillating electrons. Accordingly, the cutoff energies of high harmonics are determined by the maximum energy differences. In TDPI, the concepts of short and long trajectories in solid HHG can be clarified clearly. Furthermore, we study the CEP effects in short pulses and HHG driven by two-color fields using TDPIs. It is shown that the vector potential dominates the dynamical process of HHG in solids. The TDPI method proposed in the present work provides a promising way to analyze the solid dynamics in strong field, and it would be helpful to shed light on the underlying mechanisms in future studies.

\section{Acknowledgments}
The authors thank Prof. Xue--Bin Bian for very helpful discussions. This work was supported by the National Natural Science Foundation of China under Grants No. 11234004, No. 11404123, No. 11574101, No. 11422435, and No. 11627809.

\end{document}